\documentclass[runningheads]{llncs}
\pdfoutput=1
\usepackage{xspace}
\usepackage{pgfplots}
\usepackage{comment}
\usepackage{graphicx}
\usepackage{xspace}
\usepackage{color}
\usepackage{graphicx}
\usepackage{amsmath}
\usepackage{amssymb}
\usepackage{mathrsfs}
\usepackage{epstopdf}
\usepackage{marginnote}
\usepackage{placeins}
\usepackage{enumerate}
\usepackage{tikz}
\usepackage{hyperref}
\usepackage{tikz-qtree}
\usepackage{tabularx}
\usepackage{tikz}
\usepackage{placeins}
\usetikzlibrary{decorations.pathreplacing}
\usetikzlibrary{decorations.pathreplacing,calligraphy}
\pgfplotsset{compat=1.18}

\newcommand{\length}[1]{|#1|}

\newcommand{\AND}{\mathrel{\&}}
\newcommand{\OR}{\mathrel{|}}

\newcommand{\ALGO}{HC\xspace}

\def\pp{\mathinner{\ldotp\ldotp}}

\begin{document}

\bibliographystyle{plainurl}

\title{Efficient Online String Matching through Linked Weak Factors}
\titlerunning{Efficient Online String Matching through Linked Weak Factors}

\author{Matthew N. Palmer, Simone Faro$^{\dagger}$ \and Stefano Scafiti$^{\dagger}$}
\authorrunning{M. N. Palmer, S. Faro and S. Scafiti}
\institute{
$^{\dagger}$University of Catania, Department of Mathematics and Computer Science, Italy\\
\email{\{simone.faro,stefano.scafiti\}@unict.it}\\
\email{matt.palmer@bcs.org}
}

\maketitle
\begin{abstract}
\emph{Online string matching} is a computational problem involving the search for patterns or substrings in a large text dataset, with the pattern and text being processed sequentially, without prior access to the entire text. Its relevance stems from applications in data compression, data mining, text editing, and bioinformatics, where rapid and efficient pattern matching is crucial. Various solutions have been proposed over the past few decades, employing diverse techniques. Recently, \emph{weak recognition} approaches have attracted increasing attention. This paper presents \emph{Hash Chain}, a new algorithm based on a robust weak factor recognition approach that connects adjacent factors through hashing. Despite its $O(nm)$ complexity, the algorithm exhibits a sublinear behavior in practice and achieves superior performance compared to the most effective algorithms.
\end{abstract}


\section{Introduction}
The \emph{string matching} problem consists in finding all the occurrences of a pattern $x$ of length $m$ in a text $y$ of length $n$, both strings defined over an alphabet $\Sigma$ of size $\sigma$. 
Several solutions to this problem have been proposed over the past decades~\cite{FL13}, many of which have been presented in recent years~\cite{FaroS22,FaroS23}. This suggests that interest in increasingly efficient solutions is, nevertheless, still high.

The first linear-time solution to the problem was given by Knuth, Morris and Pratt~\cite{KMP77} (KMP), whereas Boyer and Moore (BM) provided the first sub-linear solution on average~\cite{BM77}. 
The Backward-\textsc{Dawg}-Matching (BDM) algorithm~\cite{CR94} was instead the first solution to reach the optimal $\mathcal{O}(n\log_{\sigma}(m)/ m)$ time complexity on the average. 
Both the KMP and the BDM algorithms are based on an \emph{exact factor recognition}: they simulate the deterministic automaton for the language $\Sigma^{\star}x$ and the deterministic suffix automaton for the language of the suffixes of $x$, respectively. Many of the papers in the literature have amply demonstrated (see for instance~\cite{PT03,DPST10,CFG12,FL12}) how the efficiency of these solutions is strongly affected by the encoding used for simulating the underlying automaton, especially when
the bit-parallelism technique~\cite{BYG92} is used.  For this reason, recent research~\cite{Faro16,FaroS22,CFP17,DPST10} has focused more on approaches based on a \emph{weak recognition}.

\subsubsection{Weak Recognition} We say that a structure performs a \emph{weak recognition} when it is able to recognize a broader language than the one formed by just the pattern sub-strings. 
The Backward Oracle Matching algorithm \cite{AllauzenCR99} (BOM) can be considered the pioneer of this approach, which makes use of 
the Factor Oracle of the reverse pattern. The Weak Factor Recognition algorithm~\cite{CFP17} (WFR) approach is based on indexing all the $O(m^2)$ subsequences of the pattern $x$ using a \emph{bloom filter} \cite{BLOOM70}. In ~\cite{DPST10}, Q-gram Filtering (QF) ensures that $q$-grams read in a window all belong to the same chain of $q$-grams in the pattern.  Faro and Scafiti later introduced the Range Automaton~\cite{FaroS22}, a non-standard, weak version of the non-deterministic suffix automaton.

In this paper, we introduce the Hash Chain algorithm (\ALGO), another efficient algorithm for exact string matching based on weak factor recognition and hashing. The new algorithm is based on an improved filtering approach which \emph{links} together hash values corresponding to adjacent factors of the input string $x$. 


\subsubsection{Paper organization}

The paper is organized as follows. In Section \ref{sec:Notions} we briefly introduce the basic notions which we use along the paper. 
Then in Section \ref{sec:algo} we introduce the new algorithm, describing its preprocessing and searching phase in detail.
Section \ref{sec:results} will present the results of extensive experimentation, and we draw our conclusions in Section \ref{sec:conclusions}.


\section{Basic Notions and Definitions}\label{sec:Notions}
Let $x$ be a pattern of length $m$ and $y$ a text of length $n$. Let us assume that both strings $x$ and $y$ are drawn from a common alphabet $\Sigma$ of size $\sigma$.  Given a finite alphabet $\Sigma$, we denote by $\Sigma^{m}$, with $m \geq 0$, the set of all strings of length $m$ over $\Sigma$ and put $\Sigma^{*} = \bigcup_{m\in \mathbb{N}} \Sigma^{m}$.

We represent a string $x \in \Sigma^{m}$ as an array $x[0\pp m-1]$ of characters of $\Sigma$ and write $\length{x} = m$. For $m=0$ we obtain the empty string $\varepsilon$.  
Thus, $x[i]$ is the $(i+1)$-st character of $x$, for $0\le i< m$, and $x[i\pp j]$ is the substring of $x$ contained between its $(i+1)$-st and the $(j+1)$-st characters, for $0\le i \le j < m$.

A $q$-gram is a substring of $x$ with a fixed length of $q$.  We use the following bitwise operators: OR $\OR$, AND $\AND$, and bit-shift left $\ll$.

\section{The Hash Chain Algorithm}\label{sec:algo}
We present an efficient algorithm for the exact string matching problem based on a weak-factor-recognition approach using hashing.  Our proposed algorithm is named \emph{Hash Chain} (HC) and consists of a preprocessing and a searching phase.  It finds mismatches in the text quickly by identifying $q$-grams which are not adjacent to each other in the pattern, which enables a large forward shift on average.

\subsection{The preprocessing phase}\label{sec:preprocessing}

The preprocessing phase consists of the computation of an extended bloom filter data structure indexing all the $q$-grams of the pattern $x$, each of length $q$.  It is backed by a bit-vector $F$ of $2^{\alpha}$ words, where each word has $w$ bits and $\alpha$ controls the size of $F$.  Two hash functions are used: $h: \Sigma^q \rightarrow \{0, 1, ..., 2^\alpha-1\}$, which produces an index into a word in $F$, and $\lambda: \{0, 1, ..., 2^\alpha-1\} \rightarrow \{0, 1, ..., 2^w-1\}$, which outputs a word with only one of its bits set. 

The filter is built by linking together each pair of adjacent non-overlapping factors, $u_1 \cdot u_2$ of fixed size $q$ using the following formula:

\begin{equation}\label{chain-update}
{F[h(u_2)] \leftarrow F[h(u_2)] \OR \lambda(h(u_1))}.
\end{equation}

We use the bitwise OR operator $\OR$ to retain bits already set from previous $q$-gram pairs, if they had hashed to the same word in $F$.  More formally, formula \ref{chain-update} is iterated for each pair of distinct $q$-grams $\langle u_1, u_2 \rangle$ such that:

\begin{itemize}
    \item $|u_i| = q, i = 1, 2$,
    \item $u_1 = x[i..j]$, where $j = i + q - 1$, for some $0 \leq i \leq m - 2 \cdot q$
    \item $u_2 = x[j+1..j+q]$
\end{itemize}

Note that the first $q$ positions of any pattern do not have a $q$-gram to their left. To ensure they are recognized as factors of the pattern, we index any such factor $u$ separately using the following formula:

\begin{equation}\label{first-qgrams}
{F[h(u)] \leftarrow F[h(u)] \OR 1}
\end{equation}

\begin{figure}[!t]
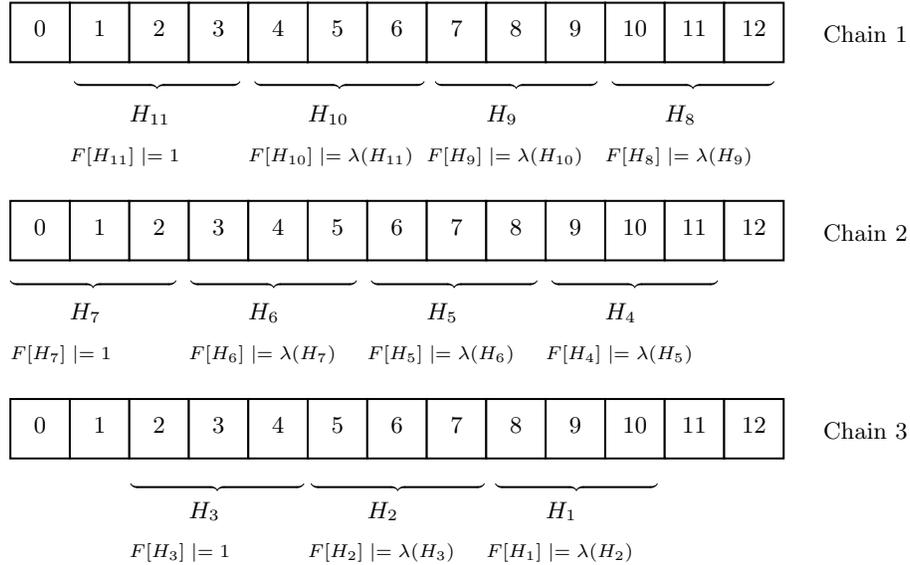

\centering
\include{hashchain-diagram}
\vspace*{-0.5cm}

\caption{The three $q$-gram chains for a pattern $x$ of length $m = 13$ and $q = 3$, and the operations performed for each $q$-gram. The $\mathrel{|}=$ symbol represents the logical \textsc{Or} operator (the result is stored in the left hand operand).  Each hash $H_n$ is labelled with a number $n$ reflecting the order in which they are calculated during pre-processing.} \label{fig:chains}
\end{figure}

To apply formula~\ref{chain-update} efficiently to each pair of non-overlapping adjacent factors of $x$, Hash Chain groups sequences of non-overlapping $q$-grams according to the position they appear in the pattern.  
More specifically, each pattern position $j$, with ${m-q \leq j < m}$, defines a sequence of $\lfloor (j+1) / q \rfloor$ non-overlapping {$q$-grams}, given by:
$$
\{x[i..i+q-1] \quad|\quad i \geq 0, i = j-q+1, j-2q+1, ...\}.
$$
Each of such sets is denoted as a \emph{$q$-gram chain} of the pattern.  By processing $q$-grams in chains, we can pass a hash value from one linked pair of $q$-grams to the computation of the next pair. This approach enables us to compute the hash only once for each $q$-gram in the pattern. Figure~\ref{fig:chains} shows the 3 chains of $q$-grams arising in a pattern of length $m = 13$ with $q = 3$.

Obviously, a pattern where $m = q$ can only have one chain, and one $q$-gram, in it.  More generally, when $m < 2 \cdot q - 1$, it only has $m - q + 1$ distinct chains, and when $m \geq 2 \cdot q - 1$, it has $q$ distinct chains of $q$-grams.

\begin{figure}[!t]
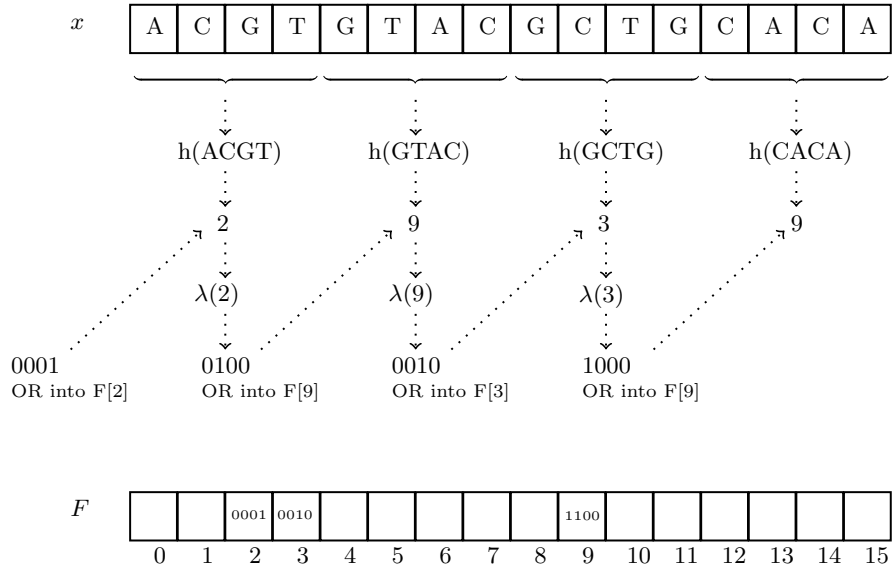

\centering
\include{qgram-linking-diagram}
\vspace*{-0.5cm}

\caption{Linking of adjacent $q$-grams $q = 4$, in a pattern $m = 16$, with $w = 4$ and $\alpha =4$.  The top part of the figure shows the calculations which are performed for the first four adjacent $q$-grams in the pattern and where they are stored.  The bottom part of the figure shows the final state of the bit vector $F$ after all calculations have been performed.  Empty cells have no entries and contain {\scriptsize 0000}.} \label{fig:qgram-linking}
\end{figure}

Figure~\ref{fig:qgram-linking} shows the process of linking four adjacent $q$-grams of the pattern ${x = \textsf{acgtgtacgctgcaca}}$.  To complete pre-processing, we would need to process the remaining chains of $q$-grams in pattern $x$: those starting at position 1 (\textsf{cgtg}, \textsf{tacg}, \textsf{ctgc}), the ones at position 2 (\textsf{gtgt}, \textsf{acgc}, \textsf{tgca}) and those at position 3 (\textsf{tgta}, \textsf{cgct} and \textsf{gcac}). 

\subsection{Hash functions}
The definitions of $h$ and $\lambda$ strongly affect the performance of the algorithm.

Function $h$ follows a \textsf{shift-then-add} structure, which can be implemented efficiently in modern architectures~\cite{NM93}. It is recursively defined as:

\begin{equation}\label{hash-function-h}
    h(x) = 
    \begin{cases}
     0   &  \text{if} \; m = 0\\
     (h(x[1..m - 1]) \cdot 2^s + x[0]) & \text{otherwise}.
    \end{cases} \mod 2^\alpha
\end{equation}

The hash value is multiplied by $2^s$, or equivalently bit-shifted left by $s$, for each additional character added. To ensure we do not shift data much beyond the maximum hash value, we calculate the bit shift $s$ by the following formula:

\begin{equation}\label{bit-shift-s}
{s \leftarrow \lfloor \alpha / q \rfloor}    
\end{equation}

The final value of the hash is taken as $\mod 2^\alpha$, which can be efficiently computed by bitwise ANDing it with $2^\alpha-1$. 

$\lambda$ is a simple function mapping each value $0 \leq v < 2^\alpha$ to the set $\{2^0, 2^1, ..., 2^{w - 1}\}$. It is meant to link together adjacent factors of the input pattern $x$, and that's why we refer to it as the link hash function.  Its definition is given by:

\begin{equation}\label{link-hash-function}
\lambda(v) = 2^{(v \mod w)}.   
\end{equation}

where $v$ is the value to obtain a link hash for and $w$ is the number of bits in a word in the bit vector $F$.  Given that $w$ is a power of two, the $\mod w$ operation can be efficiently computed by logically bitwise ANDing it with $w-1$.  It returns a word with a single bit set in it. 

\begin{figure}[!t]
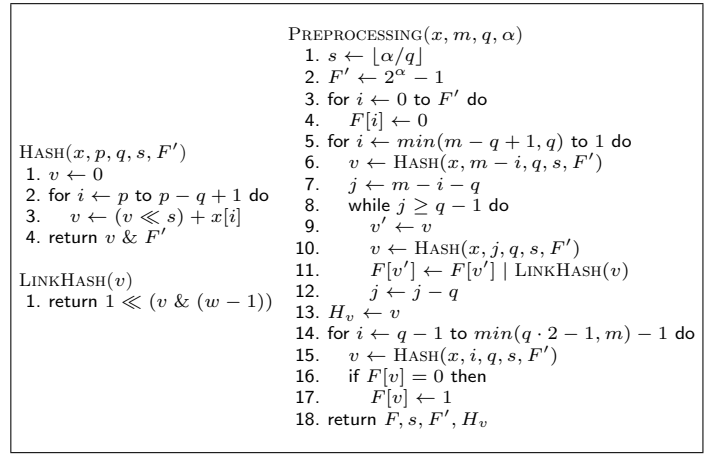

\begin{center}
\begin{scriptsize}
\begin{tabular}{|c|}
\hline
\begin{tabular}{rl}

&\\
\multicolumn{2}{l}
{\textsc{Hash($x, p, q, s, F'$)}}\\
~\textsf{1.} & \textsf{$v \leftarrow 0$}\\
~\textsf{2.} & \textsf{for $i \leftarrow p$ to $p - q + 1$ do}\\
~\textsf{3.} & \quad \textsf{$v \leftarrow (v \ll s) + x[i]$}\\
~\textsf{4.} & \textsf{return $v \AND F'$}\\

&\\
\multicolumn{2}{l}
{\textsc{LinkHash($v$)}}\\
~\textsf{1.} & \textsf{return $1 \ll (v \AND (w - 1))$}\\

&\\

\end{tabular}
\begin{tabular}{rl}
&\\

\multicolumn{2}{l}
{\textsc{Preprocessing($x, m, q, \alpha$)}}\\
~\textsf{1.} & \textsf{$s \leftarrow \lfloor \alpha / q \rfloor$}\\

~\textsf{2.} & \textsf{$F' \leftarrow 2^\alpha - 1$}\\
~\textsf{3.} & \textsf{for $i \leftarrow 0$ to $F'$ do}\\
~\textsf{4.} & \quad \textsf{$F[i]\leftarrow 0$}\\

~\textsf{5.} & \textsf{for $i \leftarrow min(m - q + 1, q)$ to $1$ do}\\
~\textsf{6.} & \quad \textsf{$v \leftarrow \textsc{Hash}(x, m - i, q, s, F')$}\\
~\textsf{7.} & \quad \textsf{$j \leftarrow m - i - q$}\\
~\textsf{8.} & \quad \textsf{while $j \geq q - 1$ do}\\
~\textsf{9.} & \quad \quad \textsf{$v' \leftarrow v$}\\
~\textsf{10.} & \quad \quad \textsf{$v \leftarrow \textsc{Hash}(x, j, q, s, F')$}\\
~\textsf{11.} & \quad \quad \textsf{$F[v'] \leftarrow F[v'] \OR \textsc{LinkHash}(v)$}\\
~\textsf{12.} & \quad \quad \textsf{$j \leftarrow j - q$}\\
~\textsf{13.} & \textsf{$H_v \leftarrow v$}\\
~\textsf{14.} & \textsf{for $i \leftarrow q - 1$ to $min(q \cdot 2 - 1, m) - 1$ do}\\
~\textsf{15.} & \quad \textsf{$v \leftarrow \textsc{Hash}(x, i, q, s, F')$}\\
~\textsf{16.} & \quad \textsf{if $F[v] = 0$ then}\\
~\textsf{17.} & \quad \quad \textsf{$F[v] \leftarrow 1$}\\
~\textsf{18.} & \textsf{return $F, s, F', H_v$}\\
&\\

\end{tabular}\\
\hline
\end{tabular}
\end{scriptsize}

\caption{\label{fig:preprocessingpseudocode}The pseudocode of the hash functions and preprocessing.}
\end{center}
\end{figure}

Pseudo code for the hash function $h$ (\textsc{Hash}), the link hash function $\lambda$ (\textsc{Link-Hash}), and the \textsc{Preprocessing} function is given in figure \ref{fig:preprocessingpseudocode}.  We don't pass $w$ into the \textsc{Link-Hash} function as it is assumed to be hard-coded.  Note that the \textsc{Preprocessing} function processes each chain of $q$-grams backwards, calculates the hashes for the first $q$ $q$-grams last, and returns a hash value $H_v$ in order to facilitate some optimisations discussed in section \ref{sec:optimisations}.  It uses $min()$ functions on lines 5 and 14 to ensure that we only process $q$-gram chains that actually exist in the pattern, as short patterns where $m < 2 \cdot q - 1$ have fewer than $q$ chains. 

Regarding the complexity of the preprocessing phase, the time is proportional to the number of $q$-grams in the pattern.  A pattern contains $m - q + 1$ distinct $q$-grams in it, each of which requires a hash computing for it once if the $q$-grams are processed in chains.  It also re-computes the hash for the first $q$ q-grams again to set a bit for them.  Each $q$-gram requires $O(q)$ time to be read, so the complexity is $O(m \cdot q)$.

\subsection{The Searching Phase}\label{sec:searching}

The searching phase works like any factor algorithm, such as BOM, WFR or QF.  The difference between them lies in how valid factors of the pattern are determined.  

\begin{figure}[!t]
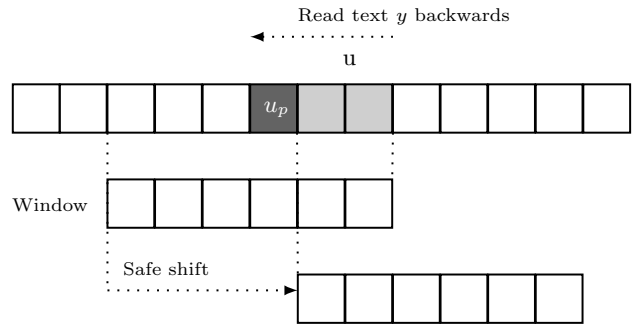

\centering
\include{factor-based-searching-diagram}
\vspace*{-0.5cm}
\caption{The general approach of factor based search algorithms.  A factor $u$ is read backwards from the end of the current window until $u_p u$ is not a factor of the pattern; it is then safe to shift the window past $u_p$.  If the entire window is read, then a possible match must be verified when a weak recognition approach is used.} \label{fig:factor-based}
\end{figure}

A window of size $m$ is slid along the text, starting at position 0, and shifted to the right after each attempt, until we reach the end of the text.  A factor $u$ of the pattern is read backwards in the text $y$, from the position aligned with the end of the window. If $u_p u$ is not a factor of the pattern, then it is safe to shift the window after $u_p$.  This is shown in figure \ref{fig:factor-based}.  

 The pseudocode of the \ALGO search algorithm is given in figure~\ref{fig:searchpseudocode}.  We determine whether a $q$-gram is a valid factor $u$ by first calculating its hash value $v$.  For the first $q$-gram processed, aligned with the end of the current window, it looks in the word at $F[v]$ to see if it is empty at line 6.  If a word in $F$ is empty, no $q$-gram in the pattern hashed to $v$, and so $u$ cannot be a factor.  We then shift on from the end of the window by the maximum distance it is possible to shift without missing a possible occurrence, $m - q + 1$, and look at the next window.  

If $F[v] \neq 0$, then we have a possible match for $u$ and the algorithm enters the \textsf{while else} loop at line 8. Hash chain must now look in turn at the $q$-grams in the window preceding $u$ to see if they are also possible factors of the pattern.  Since pre-processing links adjacent $q$-grams together with the $\lambda$ function, we calculate the hash value $v_p$ of the preceding factor $u_p$, and check to see if the bit returned by $\lambda(v_p)$ is set in $F[v]$ at line 11.  If the bit is not set, then the two factors were not linked during pre-processing, and $u_pu$ cannot be a factor of the pattern.  It is then safe to shift from the position of $u_p$ and look at the next window.

This is repeated until we reach the start of the current window, when the else branch at line 14 executes.   On line 15, the position of the search $j$ is updated such that adding $j + m - q + 1$ to it on line 18 results in the window being advanced only by one.  Finally, a check for the existence of the pattern is executed from line 16 to verify the actual presence of the pattern before reporting any occurrence. Note that, according to the \textsf{while else} semantics., the else branch at line 14 is only taken if the execution of the loop has not been interrupted by the \textsf{break} statement at line 12: naive check is thus not executed unless the window has been fully scanned.

The window is eventually advanced at line 18 depending on the size of the scanned window, thus starting a new iteration of the algorithm.

The searching phase has an $O(nm)$ complexity in the worst case and requires $O(2^\alpha)$ additional space. 

\begin{figure}[!t]
\begin{center}
\begin{scriptsize}
\begin{tabular}{|c|}
\hline
\begin{tabular}{rl}

&\\
\multicolumn{2}{l}
{\textsc{\ALGO($x, m, y, n, q, \alpha$)}}\\
~\textsf{1.} & \textsf{$F, s, F', H_v \leftarrow \textsc{Preprocessing}(x, m, q, \alpha)$}\\
~\textsf{2.} & \textsf{$j \leftarrow m-1$}\\
~\textsf{3.} & \textsf{while $j < n$ do}\\

~\textsf{4.} & \quad \textsf{$v \leftarrow \textsc{Hash}(y, j, q, s, F')$}\\
~\textsf{5.} & \quad \textsf{$z \leftarrow F[v]$}\\
~\textsf{6.} & \quad \textsf{if $z \neq 0$ then}\\

~\textsf{7.} & \quad \quad \textsf{$i \leftarrow j - m + 2 \cdot q$}\\
~\textsf{8.} & \quad \quad \textsf{while $j \geq i$ do}~~\\

~\textsf{9.} & \quad \quad \quad \textsf{$j \leftarrow j - q$}\\
~\textsf{10.} & \quad \quad \quad \textsf{$v \leftarrow \textsc{Hash}(y, j, q, s, F'$)}\\
~\textsf{11.} & \quad \quad \quad \textsf{if $z \AND \textsc{LinkHash}(v) = 0$ then}\\
~\textsf{12.} & \quad \quad \quad \quad \textsf{break {\color{gray}(to line 18)}}\\
~\textsf{13.} & \quad \quad \quad \textsc{$z \leftarrow F[v]$}\\

~\textsf{14.} & \quad \quad \textsf{else}\\
~\textsf{15.} & \quad \quad \quad \textsf{$j \leftarrow i - q$}\\
~\textsf{16.} & \quad \quad \quad \textsf{if $v = H_v$ and $y[j-q..j-q+m-1] = x$}\\
~\textsf{17.} & \quad \quad \quad \quad \textsf{output $j - q$}\\
~\textsf{18.} & \quad \textsf{$j \leftarrow j + m - q + 1$}\\
&\\

\end{tabular}
\begin{tabular}{rl}
&\\

\multicolumn{2}{l}
{\textsc{SHC($x, m, y, n, q, \alpha$)}}\\
~\textsf{1.} & \textsf{$F, s, F', H_v \leftarrow \textsc{Preprocessing}(x, m, q, \alpha)$}\\
~\textsf{2.} & \textsf{for $i \leftarrow 0$ to $m - 1$ do}\\
~\textsf{3.} & \quad \textsf{$y[n + i] \leftarrow x[i]$}\\

~\textsf{4.} & \textsf{$j \leftarrow m-1$}\\
~\textsf{5.} & \textsf{while $j < n$ do}\\

~\textsf{6.} & \quad \textsf{while $F[\textsc{Hash}(y, j, q, s, F')] = 0$ do}\\
~\textsf{7.} & \quad \quad \textsf{$j \leftarrow j + m - q + 1$}\\

~\textsf{8.} & \quad \textsf{if $j < n$ then}\\

~\textsf{9.} & \quad \quad \textsf{$v \leftarrow \textsc{Hash}(y, j, q, s, F')$}\\
~\textsf{10.} & \quad \quad \textsf{$z \leftarrow F[v]$}\\

~\textsf{11.} & \quad \quad \textsf{$i \leftarrow j - m + 2 \cdot q$}\\
~\textsf{12} & \quad \quad \textsf{while $j \geq i$ do}~~\\

~\textsf{13.} & \quad \quad \quad \textsf{$j \leftarrow j - q$}\\
~\textsf{14.} & \quad \quad \quad \textsf{$v \leftarrow \textsc{Hash}(y, j, q, s, F'$)}\\
~\textsf{15.} & \quad \quad \quad \textsf{if $z \AND \textsc{LinkHash}(v) = 0$ then}\\
~\textsf{16.} & \quad \quad \quad \quad \textsf{break {\color{gray}(to line 22)}}\\
~\textsf{17.} & \quad \quad \quad \textsc{$z \leftarrow F[v]$}\\

~\textsf{18.} & \quad \quad \textsf{else}\\
~\textsf{19.} & \quad\quad \quad \textsf{$j \leftarrow i - q$}\\
~\textsf{20.} & \quad\quad \quad \textsf{if $v = H_v$ and $y[j-q..j-q+m-1] = x$}\\
~\textsf{21.} & \quad\quad \quad \quad \textsf{output $j - q$}\\
~\textsf{22.} & \quad \quad \textsf{$j \leftarrow j + m - q + 1$}\\
&\\

\end{tabular}\\
\hline
\end{tabular}
\end{scriptsize}

\caption{\label{fig:searchpseudocode}The pseudocode of the \ALGO and SHC search algorithms.}
\end{center}
\end{figure}



\subsection{Optimisations}\label{sec:optimisations}
Several optimisations of Hash Chain are implemented in the bench-marked versions.  The first two optimisations are included in the basic Hash Chain (HC) algorithm, and the Sentinel Hash Chain algorithm (SHC) additionally implements the last one.

\begin{itemize}
    \item Reduce the bits set for the first $q$ $q$-grams,
    \item Reduce need for full pattern verification,
    \item Use a sentinel pattern at the end of the text.
\end{itemize}

\subsubsection{Reduce the bits set for the first $q$ $q$-grams}
The preprocessing order originally presented in figure \ref{fig:qgram-linking} is sub-optimal in one way.  The first step was to place an arbitrary 1 bit into the hash address of the first $q$-gram, to ensure it is recognised as a factor.  This made the process easier to describe, as we simply started at the start and proceeded to the end.

However, the algorithm only requires that the word in $F$ for that $q$-gram is not \emph{empty}, in order that it can be identified as a valid factor of the pattern.  If we process all the other pairs of $q$-grams first, it is possible that a collision will occur and the entry for it will \emph{already} contain one or more bits.  In that case, there is no need to set an additional 1 into the entry, as it already flags that it is a possible factor, by not being empty.  

Therefore, when implementing pre-processing, it is advisable to process the first $q$ $q$-grams with no preceding $q$-gram \emph{last}, and to only place a bit into the entry for them if it is empty.  This strategy ensures we set as few bits as possible in the bit vector, which reduces the chance of a false positive match.  The pre-processing pseudo-code given in figure~\ref{fig:preprocessingpseudocode} already implements this optimisation in lines 14-17.

\subsubsection{Reduce need for full pattern verification}
When the algorithm reads back to the start of the current window, it always performs a full pattern verification.  However, note that if the pattern does match the current window, then the last hash value $v$ calculated on line 10 of the pseudocode in figure~\ref{fig:searchpseudocode} \emph{must} match the hash value of the first $q$-gram in the chain ending at the end of the pattern, which we will call $H_v$.  Therefore, if we return $H_v$ from the pattern pre-processing stage, we can compare $v$ with it before entering the pattern verification step at line 16.  If the hash does not match, there is no need to perform full pattern verification.

This is the reason for the ordering of pre-processing in the pseudo-code in figure~\ref{fig:preprocessingpseudocode}.  Each chain of the pattern is processed backwards from the end of the pattern.  The last chain to be processed ends at the end of the pattern.  We compute that chain back from the end of the pattern, so the last hash value computed in it is the first hash in that chain, which is $H_v$.  This can then be returned by the pre-processing stage without having to re-compute it.  The pseudo code for HC search in figure~\ref{fig:searchpseudocode} shows this optimisation on line 16, where we test that $v = H_v$ before attempting to verify that the text matches the pattern.  In figure~\ref{fig:chains}, the hash value $H_{11}$ as the first $q$-gram in the first chain would be returned as $H_v$.

\subsubsection{Use a sentinel pattern at the end of the text}
A final optimisation technique, that can be applied to many different algorithms, is the use of a sentinel pattern at the end of the text.  This technique first makes a copy of the pattern into the text, just after the end of the text to be searched, called the \emph{sentinel pattern}.  When searching, it uses a fast search loop that does not have to perform a position check.  This is because the sentinel pattern at the end of the text \emph{guarantees} we will find a copy of the pattern if we go past the end of the text, so we can safely loop without checking our position.  Once the fast loop exits, we have to check that we have not run past the end of the text, but if not, we have a possible match to consider.

This technique, while powerful, has some serious constraints for real-world use.  It requires control over the memory allocation of the text buffer to be searched, and the ability to write data into it.  Many real-world applications will not offer that control to a search algorithm, but in cases where it is possible, it can have a performance advantage.  It has been implemented and bench-marked separately as the Sentinel Hash Chain algorithm (SHC).  

Pseudo code for the SHC algorithm is given in figure~\ref{fig:searchpseudocode}.  The pattern $x$ is copied to the end of $y$ at $n$ in lines 2 and 3.  The fast loop without a position test looking for blank words in $F$ is at lines 6 and 7, and we test to see if we have run past the end of the text at $n$ in line 8.  If not, we proceed to validate the rest of the chain and the pattern as normal. 

\section{Experimental Results}\label{sec:results}
We report in this section the results of an extensive experimental comparison of the \ALGO algorithm against the most efficient solutions known in the literature for the online exact string matching problem. Specifically, the following 21 algorithms (implemented in 99 variants, depending on the values of their parameters) have been compared:

\begin{small}
\begin{itemize}
    \item AOSO$_q$: Average-Optimal variant~\cite{FKGS05} of Shift-Or~\cite{BYG92} with $2 \leq q \leq 6$;
    \item BNDM$_q$: Backward-Nondeterministic-DAWG-Matching~\cite{NR98} with $1 \leq q \leq 6$;
    \item BRAM$_q$: Backwards Range Automaton~\cite{FaroS22}, with ${3 \leq q \leq 7}$;
    \item BSDM$_q$: Backward-SNR-DAWG-Matching~\cite{FL12}, with $2 \leq q \leq 8$;
    \item BSX$_q$: Backward-Nondeterministic-DAWG~\cite{DPST10}, with $1 \leq q \leq 8$;
    \item EBOM: Extended version~\cite{Faro2009EfficientVO} of BOM;
    \item FJS algorithm ~\cite{Franek2005ASF};
    \item LBNDM: Long BNDM algorithm~\cite{PT03};
    \item KBNDM: Factorized BNDM algorithm~\cite{CFG12};
    \item FSBNDM$_{q,s}$: Forward Simplified~\cite{CFG12} BNDM~\cite{NR98}, with $2 \leq q \leq 8$ and $1 \leq s \leq 6$;
    \item HASH$_q$: Hashing algorithm~\cite{Lecroq2007}, with $3 \leq q \leq 8$;
    \item HC$_{q,\alpha}$: Hash Chain, and its variant SHC$_{q,\alpha}$, with $1 \leq q \leq 8$ and $8 \leq \alpha \leq 12$.
    \item IOM and WOM: Improved Occurrence and Worst Occurrence Matching~\cite{Cantone2014ImprovedAS};
    \item QF$_{q,s}$: Qgram-Filtering algorithm~\cite{DPST10}, with $2 \leq q \leq 16$ and $1 \leq s \leq 6$;
    \item SBNDM$_q$: Simplified BNDM~\cite{SBNDM2009} with $2 \leq q \leq 8$;
    \item WFR$_q$: Weak Factor Recognition~\cite{CFP19}, with $1 \leq q \leq 8$ and its variant TWFR$_q$;
    \item UFM$_q$: Unique Factor Matcher~\cite{ScafitiFaro2021}, with $1 \leq q \leq 10$.
\end{itemize}
\end{small}

For completeness, we also included the Exact Packed String Matching (EPSM) algorithm~\cite{FO14}, which can only report counts but not the positions of occurrences.  Although we report its timings, we do not compare it with the other algorithms.

All algorithms have been implemented in the \textsf{C} programming language and have been tested using the \textsc{Smart} tool~\cite{FL11}. All experiments have been executed locally on a computer running Linux Ubuntu 22.04.1 with an Intel
Xeon E3-1226 v3 CPU @ 3.30GHz and 24GB ECC RAM.\footnote{The source code for the new algorithm and the \textsc{Smart} tool are available for download respectively at \url{https://github.com/nishihatapalmer/HashChain} and \url{https://github.com/smart-tool/smart}.}

Our tests have been run on a genome sequence, a protein sequence, and an English text (each of size 100MB) extracted from the well known \emph{Pizza}\&\emph{Chilli Corpus}\footnote{The corpus is available at \url{http://pizzachili.dcc.uchile.cl/index.html}.}.
In the experimental evaluation, patterns of length $m$ were randomly extracted from the sequences, with $m$ ranging over the set of values ${\{2^i\ |\ 3 \leq i \leq 10\}}$. In all cases, the mean over the search speed plus the pre-processing time (expressed in milliseconds) of $500$ runs for each pattern length has been reported.

Tables \ref{table:genomeresults}, \ref{table:proteinresults} and \ref{table:englishresults} summarise our evaluations. Each table is divided into five blocks. The first block contains algorithms based on automata.  The second contains algorithms based on character comparison.  The third block contains algorithms which use weak factor recognition, which includes the Hash Chain algorithm.  The fourth block contains algorithms that modify the text buffer to use a "sentinel" optimisation technique; all of these are also weak factor algorithms.  The final block contains algorithms which are limited to only reporting a count of occurrences, but not their positions.

Results within 105\% of the best time are underlined, and best results have been boldfaced (without considering EPSM in the final block).  For algorithms with variant parameters, such as the $q$-gram length, only the fastest variant is presented in brackets in a subscript next to the result.

\newcommand{\best}[1]{\textbf{\underline{#1}}}
\newcommand{\param}[1]{{\tiny $_{(#1)}$}}
\newcommand{\near}[1]{\underline{#1}}
\renewcommand{\arraystretch}{1.2} 

\begin{table}[!t]
\begin{center}
\begin{scriptsize}

\rotatebox[origin=c]{90}{\textsc{Genome Sequence}}~~\begin{tabular*}{1.05\textwidth}{@{\extracolsep{\fill}}|l|llllllll|}
\hline
\emph{m} & 8 & 16 & 32 & 64 & 128 & 256 & 512 & 1024 \\
\hline
\textsc{AOSO$_q$} & 63.17\param{2} & 38.65\param{4} & 19.66\param{6} & 19.75\param{6} & 19.71\param{6} & 19.74\param{6} & 19.66\param{6} & 19.71\param{6} \\
\textsc{BNDM$_q$} & 37.48\param{4} & 19.63\param{4} & 10.35\param{6} & 10.24\param{6} & 10.28\param{6} & 10.26\param{6} & 10.19\param{6} & 10.22\param{6} \\
\textsc{\textbf{BSDM$_q$}} & \best{29.02}\param{4} & 15.59\param{6} & 9.3\param{7} & \near{7.45}\param{8} & \near{7.19}\param{8} & 7.27\param{7} & 7.34\param{7} & 7.39\param{6} \\
\textsc{BXS$_q$} & 37.06\param{4} & 19.04\param{4} & 9.72\param{6} & 9.74\param{6} & 9.71\param{6} & 9.73\param{6} & 9.71\param{6} & 9.75\param{6} \\
\textsc{EBOM} & 95.02 & 64.79 & 40.09 & 24.51 & 14.58 & 9.63 & 7.75 & 4.97 \\
\textsc{FSBNDM$_{q,s}$} & 35.11\param{4,1} & 19.13\param{6,2} & 10.42\param{6,1} & 10.38\param{6,1} & 10.47\param{6,1} & 10.49\param{6,1} & 10.35\param{6,1} & 10.39\param{6,1} \\
\textsc{KBNDM} & 107.38 & 68.98 & 39.31 & 23.83 & 23.2 & 23.22 & 23.19 & 23.35 \\
\textsc{LBNDM} & 140.1 & 80.12 & 45.44 & 36.13 & 30.05 & $>$200 & $>$200 & $>$200 \\
\textsc{SBNDM$_q$} & 36.84\param{4} & 19.03\param{4} & 10.5\param{6} & 10.52\param{6} & 10.49\param{6} & 10.49\param{6} & 10.4\param{6} & 10.41\param{6} \\

\hline
\textsc{FJS} & $>$200 & $>$200 & $>$200 & $>$200 & $>$200 & $>$200 & $>$200 & $>$200 \\
\textsc{HASH$_q$} & 88.03\param{3} & 42.34\param{3} & 21.85\param{5} & 12.86\param{5} & 11.12\param{5} & 11.68\param{5} & 13.73\param{5} & 12.41\param{5} \\
\textsc{IOM} & $>$200 & $>$200 & $>$200 & $>$200 & $>$200 & $>$200 & $>$200 & $>$200 \\
\textsc{WOM} & $>$200 & $>$200 & $>$200 & 97.49 & 84.66 & 72.05 & 64.32 & 56.83 \\

\hline
\textsc{BRAM$_q$} & 58.56\param{5} & 25.13\param{5} & 12.67\param{7} & 8.76\param{7} & 7.83\param{7} & 6.92\param{7} & 4.03\param{7} & 2.95\param{7} \\
\textsc{\textbf{HC$_{q,\alpha}$}} & \near{30.32}\param{4,12} & 14.2\param{6,12} & \best{8.54}\param{6,12} & \best{7.14}\param{6,12} & \near{7.06}\param{6,12} & \near{5.75}\param{6,12} & 3.35\param{6,12} & 2.13\param{8,12} \\
\textsc{QF$_{q,s}$} & 33.39\param{4,3} & 14.69\param{4,3} & \near{8.66}\param{6,2} & \near{7.39}\param{6,2} & \near{7.08}\param{6,2} & \near{5.77}\param{6,2} & 3.43\param{6,2} & 2.52\param{6,2} \\
\textsc{UFM{$_q$}} & 42.0\param{5} & 18.38\param{6} & 9.96\param{7} & 7.71\param{8} & 7.68\param{8} & 6.46\param{8} & 3.66\param{8} & 2.22\param{1,0} \\
\textsc{WFR${_q}$} & 35.62\param{4} & 16.63\param{5} & 9.93\param{5} & 7.72\param{7} & \near{7.17}\param{6} & \near{5.79}\param{6} & \near{3.25}\param{7} & \near{2.08}\param{7} \\

\hline
\textsc{\textbf{SHC$_{q,\alpha}$}} & \near{29.65}\param{4,12} & \best{12.82}\param{5,12} & \near{8.68}\param{6,12} & \near{7.39}\param{6,12} & \best{7.03}\param{6,12} & \best{5.63}\param{6,12} & \near{3.29}\param{6,12} & \near{2.09}\param{8,12} \\
\textsc{\textbf{TWFR${_q}$}} & 31.04\param{4} & 15.68\param{5} & 9.33\param{6} & 7.62\param{6} & \near{7.08}\param{6} & \near{5.66}\param{6} & \best{3.17}\param{7} & \best{2.02}\param{7} \\

\hline
\textsc{EPSM} & 22.93 & 9.96 & 6.82 & 6.63 & 5.57 & 3.5 & 1.98 & 1.32 \\

\hline
\end{tabular*}\\[0.2cm]

\end{scriptsize}
\end{center}
\caption{Experimental results obtained for searching on a genome sequence.  Entries that state $>200$ indicate that the algorithm took longer than 200ms to search and timed out.}
\label{table:genomeresults}
\end{table}


\begin{table}[!t]
\begin{center}
\begin{scriptsize}

\rotatebox[origin=c]{90}{\textsc{Protein Sequence}}~~\begin{tabular*}{1.05\textwidth}{@{\extracolsep{\fill}}|l|llllllll|}
\hline
\emph{m} & 8 & 16 & 32 & 64 & 128 & 256 & 512 & 1024 \\
\hline
\textsc{AOSO$_q$} & 33.7\param{4} & 24.18\param{4} & 16.35\param{6} & 16.17\param{6} & 16.16\param{6} & 16.16\param{6} & 16.2\param{6} & 16.17\param{6} \\
\textsc{BNDM$_q$} & 19.18\param{2} & 11.96\param{2} & 8.36\param{4} & 8.17\param{4} & 8.16\param{4} & 8.18\param{4} & 8.19\param{4} & 8.13\param{4} \\
\textsc{\textbf{BSDM$_q$}} & 17.63\param{3} & 10.06\param{4} & \near{7.58}\param{4} & \best{6.8}\param{4} & 6.69\param{4} & 6.66\param{4} & 6.62\param{4} & 6.55\param{4} \\
\textsc{\textbf{BXS$_q$}} & \best{15.51}\param{2} & 9.95\param{3} & \near{7.77}\param{4} & 7.76\param{4} & 7.76\param{4} & 7.77\param{3} & 7.79\param{4} & 7.77\param{4} \\
\textsc{EBOM} & \near{15.98} & 10.97 & 8.97 & 8.11 & 7.06 & 5.52 & 3.11 & 1.99 \\
\textsc{FSBNDM$_{q,s}$} & \near{15.87}\param{2,0} & 9.96\param{3,1} & 7.96\param{3,1} & 7.98\param{3,1} & 7.97\param{3,1} & 7.98\param{3,1} & 7.95\param{3,1} & 7.98\param{3,1} \\
\textsc{KBNDM} & 45.47 & 25.63 & 14.9 & 11.95 & 10.95 & 11.42 & 11.41 & 11.37 \\
\textsc{LBNDM} & 68.43 & 42.74 & 20.28 & 14.2 & 11.8 & 9.89 & 9.21 & 12.56 \\
\textsc{SBNDM$_q$} & \near{15.83}\param{2} & 10.85\param{2} & 8.36\param{4} & 8.32\param{4} & 8.33\param{4} & 8.33\param{4} & 8.33\param{4} & 8.33\param{4} \\

\hline
\textsc{FJS} & 69.82 & 46.86 & 35.53 & 31.05 & 28.51 & 27.39 & 27.22 & 26.64 \\
\textsc{HASH$_q$} & 80.9\param{3} & 37.23\param{3} & 19.5\param{3} & 12.55\param{5} & 10.79\param{5} & 11.48\param{3} & 12.9\param{3} & 12.36\param{3} \\
\textsc{IOM} & 62.5 & 41.85 & 31.74 & 27.41 & 25.23 & 24.26 & 24.22 & 23.8 \\
\textsc{WOM} & 67.28 & 43.7 & 31.41 & 25.62 & 21.99 & 19.92 & 18.49 & 17.89 \\

\hline
\textsc{BRAM$_q$} & 31.48\param{3} & 16.36\param{3} & 11.15\param{3} & 8.58\param{7} & 7.79\param{7} & 6.81\param{7} & 3.71\param{7} & 2.42\param{7} \\
\textsc{HC$_{q,\alpha}$} & \near{16.17}\param{3,11} & \near{9.38}\param{3,11} & \near{7.58}\param{3,11} & \near{6.84}\param{6,12} & \near{6.23}\param{3,11} & \near{4.12}\param{3,11} & \near{2.37}\param{4,12} & \near{1.42}\param{4,12} \\
\textsc{QF$_{q,s}$} & \near{16.03}\param{2,6} & 9.93\param{3,4} & \near{7.63}\param{3,4} & \near{6.84}\param{4,3} & \near{6.27}\param{3,4} & \near{4.16}\param{3,4} & \near{2.33}\param{4,3} & \near{1.4}\param{4,3} \\
\textsc{UFM$_q$} & 23.15\param{3} & 13.66\param{3} & 9.69\param{7} & 7.72\param{8} & 7.68\param{8} & 6.37\param{7} & 3.54\param{8} & 2.17\param{1,0} \\
\textsc{WFR$_q$} & 26.03\param{2} & 12.36\param{4} & 8.37\param{4} & 7.2\param{4} & 6.61\param{4} & 4.75\param{4} & 2.52\param{5} & 1.5\param{5} \\

\hline
\textsc{\textbf{SHC$_{q,\alpha}$}} & \near{15.85}\param{3,11} & \best{9.23}\param{3,11} & \best{7.49}\param{3,11} & \near{6.82}\param{4,12} & \best{6.19}\param{3,11} & \best{4.06}\param{3,11} & \best{2.29}\param{4,12} & \best{1.37}\param{4,12} \\
\textsc{TWFR$_q$} & 23.37\param{4} & 10.73\param{4} & 8.09\param{4} & \near{7.1}\param{4} & 6.52\param{4} & 4.62\param{4} & 2.49\param{5} & 1.48\param{5} \\

\hline
\textsc{EPSM} & 11.44 & 10.06 & 6.87 & 6.69 & 5.62 & 3.52 & 1.95 & 1.34 \\

\hline
\end{tabular*}\\[0.2cm]

\end{scriptsize}
\end{center}
\caption{Experimental results obtained for searching on a protein sequence. }
\label{table:proteinresults}
\end{table}

\begin{table}[!t]
\begin{center}
\begin{scriptsize}

\rotatebox[origin=c]{90}{\textsc{English Text}}~~\begin{tabular*}{1.05\textwidth}{@{\extracolsep{\fill}}|l|llllllll|}
\hline
\emph{m} & 8 & 16 & 32 & 64 & 128 & 256 & 512 & 1024 \\
\hline
\textsc{AOSO$_q$} & 36.67\param{4} & 24.19\param{4} & 16.3\param{6} & 16.11\param{6} & 16.11\param{6} & 16.08\param{6} & 16.09\param{6} & 16.1\param{6} \\
\textsc{BNDM$_q$} & 24.55\param{2} & 13.37\param{4} & 8.9\param{4} & 8.83\param{4} & 8.79\param{4} & 8.83\param{4} & 8.76\param{4} & 8.86\param{4} \\
\textsc{BSDM$_q$} & 18.92\param{3} & \near{10.8}\param{4} & \near{8.07}\param{4} & \near{7.07}\param{6} & 6.85\param{6} & 6.87\param{4} & 6.81\param{4} & 6.16\param{8} \\
\textsc{BXS$_q$} & 21.18\param{2} & 11.92\param{3} & \near{8.33}\param{4} & 8.31\param{4} & 8.31\param{4} & 8.36\param{4} & 8.29\param{4} & 8.34\param{4} \\
\textsc{EBOM} & 21.43 & 16.36 & 13.71 & 11.78 & 9.41 & 7.13 & 4.62 & 3.16 \\
\textsc{FSBNDM$_{q,s}$} & 20.44\param{3,1} & 11.97\param{3,1} & 8.71\param{4,2} & 8.71\param{4,2} & 8.7\param{4,2} & 8.75\param{4,2} & 8.68\param{4,2} & 8.7\param{4,2} \\
\textsc{KBNDM} & 49.08 & 30.41 & 19.88 & 15.23 & 11.89 & 11.82 & 11.81 & 11.85 \\
\textsc{LBNDM} & 81.15 & 49.51 & 26.67 & 16.95 & 13.08 & 10.47 & 9.13 & 10.72 \\
\textsc{SBNDM$_q$} & 20.75\param{2} & 13.41\param{4} & 8.89\param{4} & 8.84\param{4} & 8.83\param{4} & 8.9\param{4} & 8.83\param{4} & 8.88\param{4} \\

\hline
\textsc{FJS} & 75.05 & 49.48 & 36.19 & 29.83 & 24.97 & 21.69 & 20.46 & 18.29 \\
\textsc{HASH$_q$} & 80.36\param{3} & 36.76\param{3} & 18.67\param{3} & 12.33\param{3} & 10.71\param{3} & 11.49\param{3} & 13.87\param{3} & 12.18\param{3} \\
\textsc{IOM} & 67.84 & 45.18 & 33.21 & 26.8 & 22.51 & 19.83 & 19.18 & 17.38 \\
\textsc{WOM} & 72.35 & 45.86 & 31.92 & 25.4 & 20.48 & 17.87 & 16.75 & 16.51 \\

\hline
\textsc{BRAM$_q$} & 31.99\param{3} & 17.98\param{3} & 10.85\param{5} & 8.35\param{5} & 7.7\param{5} & 6.14\param{5} & 3.23\param{7} & 1.83\param{7} \\
\textsc{\textbf{HC$_{q,\alpha}$}} & \near{17.54}\param{3,11} & \near{10.67}\param{3,11} & \near{8.08}\param{6,12} & \best{6.84}\param{6,12} & \near{6.49}\param{3,11} & \near{4.57}\param{3,11} & \near{2.58}\param{4,12} & 1.58\param{4,12} \\
\textsc{QF$_{q,s}$} & 20.28\param{3,4} & \near{10.47}\param{4,3} & \near{8.06}\param{4,3} & \near{6.99}\param{4,3} & \near{6.5}\param{4,3} & \near{4.64}\param{4,3} & 2.61\param{4,3} & 1.69\param{4,3} \\
\textsc{UFM$_q$} & 24.71\param{3} & 13.94\param{5} & 9.02\param{6} & 7.47\param{6} & 7.4\param{6} & 5.69\param{6} & 3.21\param{8} & 1.99\param{8} \\
\textsc{WFR$_q$} & 26.47\param{2} & 12.75\param{4} & 8.67\param{4} & 7.3\param{4} & \near{6.63}\param{4} & 4.77\param{5} & \near{2.54}\param{5} & \near{1.5}\param{5} \\

\hline
\textsc{\textbf{SHC$_{q,\alpha}$}} & \best{17.14}\param{3,11} & \best{10.41}\param{3,11} & \best{8.0}\param{5,12} & \near{7.0}\param{5,12} & \best{6.44}\param{3,11} & \best{4.52}\param{3,11} & \best{2.48}\param{4,12} & \near{1.55}\param{4,12} \\
\textsc{\textbf{TWFR$_q$}} & 23.52\param{4} & 11.14\param{4} & \near{8.31}\param{4} & \near{7.13}\param{4} & \near{6.53}\param{4} & \near{4.67}\param{4} & \near{2.51}\param{5} & \best{1.49}\param{5} \\

\hline
\textsc{EPSM} & 12.83 & 9.95 & 6.81 & 6.62 & 5.56 & 3.47 & 1.94 & 1.28 \\

\hline
\end{tabular*}\\[0.2cm]

\end{scriptsize}
\end{center}
\caption{Experimental results obtained for searching on an English text.}
\label{table:englishresults}
\end{table}


For all alphabets and pattern lengths $m > 8$, HC$_{q,\alpha}$ and its variant SHC$_{q,\alpha}$ are almost always fastest.  On protein sequences only, BSDM$_q$ achieves a fastest time when $m = 64$.  On genome sequences TWFR$_q$ is fastest when $m \geq 512$ and on English when $m = 1024$.  With the exception of the BSDM result, \emph{all} of the fastest algorithms with $m \geq 16$ are based on the weak factor recognition approach, while for $m < 16$, the fastest algorithms are almost always automata-based.  

Comparing SHC and HC, we can see that SHC is the faster of the two variants, as it achieves the greatest number of fastest results over all alphabets, and with the exception of genome is almost always the fastest algorithm for all pattern lengths.  In most of the cases where SHC is not the fastest, HC is the fastest.  Where either algorithm is not the fastest, they are almost always within 105\% of the fastest time. 

\section{Conclusions}\label{sec:conclusions}
In this paper we introduced the Hash Chain algorithm (\ALGO) and its variant SHC, a new exact string matching algorithm based on weak factor recognition and hashing which links adjacent hash values of the pattern $x$.
From extensive experimental evaluation, our newly presented algorithm is extremely competitive when compared with the most efficient algorithms known in the literature.

The good performance obtained by the \ALGO and SHC algorithms, and other similar weak factor recognizers, suggest that weak factor recognition is a  promising approach in the field of pattern recognition, encouraging further research on the same direction.  Moreover, a linear version of \ALGO and SHC should be possible, modelled on the same lines as the Linear Weak Factor Recognition algorithm\cite{CFP19}.

\FloatBarrier
\begin{scriptsize}
\bibliography{main}
\end{scriptsize}

\end{document}